\documentclass{aa}
\usepackage{graphics,epsf,epsfig,graphicx,latexsym}

\begin{document}

 \topmargin .3cm

\title{Kinematics of tidal tails in interacting galaxies:\\
tidal dwarf galaxies and projection effects\thanks{Based on observations collected
at the European Southern Observatory, La Silla, Chile and at the
Canada-France-Hawaii Observatory, Hawaii, USA.}}
                
\author{F. Bournaud \inst{1,2,3}, P.-A. Duc \inst{1,4}, P. Amram \inst{5}, F. Combes \inst{2}, and J.-L. Gach \inst{5}}
 
\offprints{F. Bournaud \email{Frederic.Bournaud@obspm.fr}} 
\institute{
CEA/DSM/DAPNIA, Service d'Astrophysique, Saclay, 91191 Gif sur Yvette Cedex, France
\and
Observatoire de Paris, LERMA, 61 Av. de l'Observatoire, F-75014, Paris, France
\and
Ecole Normale Sup\'erieure, 45 rue d'Ulm, F-75005, Paris, France
\and
CNRS FRE 2591
\and
Observatoire Astronomique Marseille-Provence \& Laboratoire d'Astrophysique de Marseille, 2 Place Le Verrier, 13248 Marseille Cedex 04, France}
\date{Received XX XX, 2004; accepted XX XX, 2004}
\authorrunning{Bournaud et al.} 
\titlerunning{Kinematics of tidal tails and tidal dwarf galaxies}

\abstract{The kinematics of tidal tails in colliding galaxies has been 
studied via Fabry-P\'erot observations of the H$\alpha$  emission. With 
their large field of view and high spatial resolution, the Fabry-P\'erot 
data allow us to probe simultaneously, in 2-D, two kinematical features of 
the tidal ionized gas: large-scale velocity gradients due to streaming 
motions along the tails, and small-scale motions related to the internal 
dynamics of giant HII regions within the tails. In several interacting 
systems, massive (10$^9$ M$_{\sun}$) condensations of HI, CO and stars are 
observed in the outer regions of tails. Whether they are genuine 
accumulations of matter or not is still debated. Indeed a part of the 
tidal tail may be aligned with the line-of-sight, and the associated 
projection effect may result in apparent accumulations of matter that does 
not exist in the 3-D space. Using numerical simulations, we show that 
studying the large-scale kinematics of tails, it is possible to know 
whether these accumulations of matter are the result of projection effects 
or not. We conclude that several ones (Arp~105--South, Arp~242, NGC~7252, 
and NGC~5291--North) are genuine accumulations of matter. We also study 
the small-scale motions inside these regions: several small-scale velocity 
gradients are identified with projected values as large as 50--100 
km~s$^{-1}$ accross the observed HII regions. In the 
case of NGC~5291--North, the spatial resolution of our observations is 
sufficient to detail the velocity field; we  show that this system is 
rotating and self-gravitating, and discuss its dark matter content. The 
Fabry-P\'erot observations have thus enabled us to prove that some 10$^9$ 
M$_{\sun}$ condensations of matter are real structures, and are kinematically decoupled from the rest of the tail. Such massive and self-gravitating objects are the progenitors of the so-called ''Tidal Dwarf Galaxies''.
}

\maketitle

%%%%%%%%%%%%%%%%%%%%%%%%%%%%%%%%%%%%%%%%%%%%%%%%%%%%%%%%%%%%%%%%%%%%%%%%%%%%%%

\section{Introduction}

One of the most prominent features shown by interacting galaxies is their long tails made of
gas, dust and stars that were shaped by tidal forces. Their morphology has largely 
been studied in association with numerical simulations to obtain constraints on the 
age of the galactic encounter and on its initial conditions. Some properties of the tails, in 
particular their length, may also provide indirect information on the distribution of the
surrounding dark matter (Mihos et al. 1998; Dubinski et al. 1999; Bournaud, Duc \& Masset 2003).

In many interacting systems, gravitational clumps are observed along tidal tails 
 (Mirabel et al. 1992; Weilbacher et al. 2000, 2002, 2003' Iglesias-P\'aramo \& V\'ilchez 2001; Knierman et al. 2003). In fewer systems, very massive condensations of 
matter were discovered in the external regions, often near the extremity of the tails.
They have apparent masses typical of dwarf galaxies, i.e. about $10^9$ M$_{\sun}$ (Duc \& Mirabel 1994, 1998; Hibbard et al. 1994, 2001; Nordgren et al. 1997; Duc et al. 2000; Braine et al. 2001). 
Their high metallicities indicate that they are formed of pre-enriched gas and hence cannot be dwarf galaxies pre-existing to the galaxy interaction (Duc \& Mirabel 1994, 1998; Duc et al. 2000). The formation of such massive objects in the outer parts of tails can be reproduced by numerical simulations (Bournaud et al. 2003). They are usually considered as the progenitors of the so-called 
''Tidal Dwarf Galaxies'' (TDGs); yet whether they are real objects is 
still 
debated. Indeed tidal tails are not linear (see Figs.~\ref{fig4} 
and \ref{fig5} in Sect.~3), so that when a  tidal tail is seen edge-on, a part of it may be aligned with the 
line-of-sight, and an apparent mass condensation of $10^9$ M$_{\sun}$ may
result of projection effects (Hibbard \& Barnes 2004).

 Braine et al. (2000, 2001) have reported the presence of CO at the location of these HI condensations. Arguing that the formation of molecular gas out of HI implies a high local density, they claimed that they are real objects, and not the result of projection effects. However, this argument assumes that this molecular gas is not pre-existent to the formation of the tails, which has not definitively been checked.

Three fundamental questions related to the nature of the massive condensations observed in tidal tails
remain open:
\begin{enumerate}
\item Are they the result of projection effects, or are they genuine accumulations of matter?
\item If they are real objects, are they only transient accumulations of matter, or are they actually dwarf galaxies in formation?  As pointed out by Duc et al. (2000) and Hibbard et al. (2001), one further needs to check whether they are kinematically decoupled from the rest of the tail and gravitationally bound.
\item Do they contain dark matter ? A positive answer would suggest a dissipative dynamics for at least a part of the dark matter. It would be a strong argument for the model of dark matter in the form of cold molecular gas (Pfenniger \& Combes 1994; Pfenniger et al. 1994). Note however that this dark molecular gas, if it exists in spirals, may be converted into neutral hydrogen, and thus be absent in these tidal condensations.
\end{enumerate}

In this paper, we study the kinematics of tidal tails and address the three questions:
the role of projection effects, the dynamical properties of tidal clumps and their dark matter content. 
We have carried out optical Fabry-P\'erot (FP) observations of the ionized 
gas in several interacting systems with TDG candidates. The large field-of-view enables a direct study of the large-scale kinematics, in particular the streaming motions along the tails, while their high spatial resolution gives access to the inner dynamics of the tidal objects. Unfortunately, the HII regions in tidal tails are generally sparsely distributed. We have hence selected  systems in which the H$\alpha$ emission is observed along the whole tail (and not only in a few clumps). We present in Sect.~2 the observational techniques and the results browsing our FP data cubes. Carrying out numerical simulations of galaxy collisions, we found a signature of projection effects based on the  large-scale kinematics of the tails. In Sect.~3, we investigate them in the systems for which we have obtained FP data and for a few objects with interferometric HI data as well. In Sect.~4 , we study the small-scale kinematics of tidal condensations. Strong velocity gradients inside several TDG candidates had previously been found in long-slit spectroscopy observations and interpreted as a signature of self-gravity and of a kinematical decoupling (e.g., Duc et al. 1997; Duc \& Mirabel 1998; Weilbacher et al. 2002). Our new Fabry-P\'erot observations 
were used to confirm them, and provide the direction of the largest 
gradient in each case. The Fabry-P\'erot technique had previously been used 
with success in the study of the collisional debris of a few interacting systems (e.g., Mihos \& Bothun 1998; Mendes de Oliveira et al. 2001). Deriving from the inferred velocity curve the dynamical mass of one massive tidal condensation, we discuss its dark matter content. The main conclusions are drawn in Sect.~5. Observational results, H$\alpha$ emission maps and velocity fields are detailed system by system in Appendix~\ref{appen}.

%however this dark molecular gas, if it exists, may be converted into neutral then star-forming molecular gas, and thus be %absent from TDGs. 

%%%%%%%%%%%%%%%%%%%%%%%%%%%%%%%%%%%%%%%%%%%%%%%%%%%%%%%%%%%%%%%%%%%%%%%%%%%%%%

\section{Observations}

\subsection{Fabry-P\'erot observations and data reduction}

Observations were carried out during two runs at the 
Canada-France-Hawaii 3.6m telescope (CFHT 3.6m) in
February-March 2000 and at the European Southern Observatory 
3.6m telescope (ESO 3.6m) in April 2002.

At the CFHT 3.6m, the multi-object spectrograph focal reducer
(MOS), attached to the f/8 Cassegrain focus, was used in the
Fabry-P\'erot mode. The CCD was a ''fast'' STIS 2 detector, 2048
$\times$ 2048 pixels with a read-out noise of 9.3 e$^{-}$ and a
pixel size on the sky of 0.88 arcsec after 2x2 binning to increase
the signal to noise ratio

During the run at the ESO 3.6m telescope, the Fabry-P\'erot
instrument CIGALE was used. It is composed of a focal reducer
(bringing the original f/8 focal ratio of the Cassegrain focus to
f/2), a scanning Fabry-P\'erot and an Image Photon Counting System
(IPCS). The IPCS is a new generation GaAs camera. The
semi-conductor photocathode GaAs offers a high quantum efficiency. The output
of the GaAs tube is coupled by optical fibres to a 1024 x 1024 CCD
(Gach et al. 2002). The IPCS, with a time sampling of 1/50
second and zero readout noise, makes it possible to scan the
interferometer rapidly, avoiding
sky transparency, air-mass and seeing variation problems during
the exposures. Unfortunately, no IPCS was available at the CFHT.

Tables 1 and 2 contain the journal of observations and
observational characteristics for both runs. Reduction of the data
cubes was performed using the CIGALE/ADHOCw software\footnote{see http://www.oamp.fr/adhoc/adhoc.html}. The data reduction procedure has been extensively described in Amram et al. (1998) and references therein.

Wavelength calibration was obtained by scanning the narrow
spectral emission lines of a source lamp under the same conditions as the
observations. Velocities measured relative to the systemic
velocity are very accurate, with an error of a fraction of a
channel width ($< 3$~km~s$^{-1}$) over the whole field.

Subtraction of bias, flat fielding of the data and cosmic-ray
removal have been performed for each image of the data cube for
the CFHT observations. To minimize seeing variation, each scan
image was smoothed with a Gaussian function of full-width at half
maximum equal to the worst case of seeing among the scans.
Transparency and sky foreground fluctuations have also been
corrected for using field star fluxes and galaxy-free areas for the
CFHT observations. Except for flat fielding, none of these
operations are necessary for the IPCS data processing (ESO
observations).

The velocity sampling was 10 to 11 km~s$^{-1}$ at CFHT and 16 km~s$^{-1}$ at ESO. Profiles were
spatially binned to 3$\times$3 or 5$\times$5 pixels in the outer
parts, in order to increase the signal-to-noise ratio. Strong OH
night sky lines passing through the filters were subtracted by
determining the level of emission from extended regions away from
the galaxies (Amram et al. 1998).

\begin{table*}[tbp]
\centering
\begin{tabular}[t]{llll}
\hline
\hline
Observations & Telescope & CFHT 3.6m & ESO 3.6m \\
\hline
       & Equipment & MOS/FP @ Cassegrain & CIGALE @ Cassegrain\\
       & Date & 2000, Feb 28 - Mar 2 & 2002, April, 9 \\
       & Seeing & 1"-1.2" & $\sim$ 1" \\
Calibration & Comparison light & $\lambda$ 6599 \& 6717 \AA & $\lambda$ 6599 \AA \\
Fabry-P\'erot & Interference Order & 1162 @ 6562.78 \AA & 796 @ 6562.78 \AA \\
     & Free Spectral Range at H$\alpha$ & 265~km~s$^{-1}$ & 378~km~s$^{-1}$ \\
     & \textit{Finesse}$^{(1)}$ at H$\alpha$ & 14 & 12\\
     & Spectral resolution at H$\alpha$ & 13672$^{(2)}$ & 9375$^{(2)}$\\
Sampling & Sampling Step & 0.24 \AA\ (10--11~km~s$^{-1}$) & 0.35 \AA\ (16~km~s$^{-1}$)\\
   & Number of scanning Steps & 28 & 24\\
   & Elementary observing time per interferogram & 300 (secondes) & 15 (secondes)\\
   & Total Field & 352''$\Box$- 616''$\Box$ (2k px)$^2$ $^{(3)}$ & 207''$\Box$ (1k px)$^2$ \\
     & Pixel Size & 0.88''$^{(4)}$ & 0.405''$^{(4)}$ \\
Detector & & Fast STIS 2 CCD & GaAs/IPCS \\
\hline
\end{tabular}
\begin{flushleft}
$^{(1)}$ \textit{Mean \textit{Finesse} through the field of view
\\}
$^{(2)}$ \textit{For a signal to noise ratio of 5 at the sample
step
\\}
$^{(3)}$ \textit{Respectively for a raster 800 $\times $ 800 $\&$
1400 $\times $ 1400\\}
$^{(4)}$ \textit{After binning
2$\times$2\\}
\end{flushleft}
\caption[]{Journal of Fabry-P\'erot observations.} \label{table 1}
\end{table*}

\begin{table*}[tbp]
%\begin{flushleft}
\begin{tabular}[t]{llllllll}
\hline \hline

Name & Telescope & \multicolumn{2}{c}{Exposure Times} & Scanning  & \multicolumn{3}{c}{Interference Filter} \\
   &      & Total & per channel        & Wavelength &$^{(*)}$Central   & FWHM & Transmission \\
   &      &    &              & &Wavelenght &   & at maximum \\
   &      & (hour) & (second)         & (\AA)&(\AA)     &(\AA)  &      \\
\hline
NGC 5291 Center & ESO 3.6m & 1.4 & 210 &6658& 6649 & 21& 0.70 \\%#48
NGC 5291 North & ESO 3.6m & 1.2 & 180 &6650& 6649 & 21& 0.70 \\%#48
NGC 5291 South & ESO 3.6m & 1.7 & 255 &6664& 6666 & 20& 0.60 \\%#49
Arp 105 South (NGC 3561) & CFHT 3.6m & 2.3 & 300 &6756& 6758 & 30& 0.72\\ %#54
Arp 215 (NGC 2782) & CFHT 3.6m & 2.3 & 300 &6619& 6622 & 19& 0.77 \\%#4608
Arp 242 (NGC 4676) & CFHT 3.6m & 1.4 & 180 &6711& 6715 & 16& 0.70 \\%#40
Arp 243 (NGC 2623) & CFHT 3.6m & 2.3 & 300 &6684& 6689 & 23& 0.75 \\%#50
Arp 244 (NGC 4038/9) & CFHT 3.6m & 1.5 & 190 &6599& 6603 & 11& 0.69 \\%#4606
Arp 245 North & CFHT 3.6m & 1.4 & 180 &6611& 6612 & 12& 0.61 \\%#4607
Arp 245 South & CFHT 3.6m & 2.8 & 360 &6612& 6622 & 19& 0.77 \\%#4608
IC 1182 & CFHT 3.6m & 2.6 & 340 &6784& 6781 & 19& 0.68 \\%#52
\hline
\end{tabular}\\
\textit{$^{(*)}$When necessary, filters have been inclined in the
optical beam to blueshift their passband. Characteristics are given after tilt of the filter and central wavelengths have been corrected for temperature and focal beam effects.\\}
\caption[]{Observational characteristics.} \label{table 2}
\end{table*}

\subsection{Analysis of spectroscopic data cubes and 
results}\label{analyse}

The data cubes obtained after the reduction were mainly analyzed through position-velocity (PV) diagrams. We have developed numerical tools that enable us to get:
\begin{itemize}
\item the usual PV diagram along a linear slit
%\item a PV diagram along a non linear slit, in order to precisely follow 
the large-scale morphology of the tails
\item a PV diagram along a band instead of a thin slit: if $x$ is the position along the slit, $y$ the position perpendicular to it, $v$ the velocity, and $I(v,x,y)$ the intensity in the data cube, we represent $\int_y I(v,x,y) \mathrm{d}y$ instead of $I(v,x,y=0)$, where the integral is computed on a chosen number of pixels, usually the width of the tail. This integration over the width of the tail not only reduces the noise, but also includes a larger number of HII regions. Hence, kinematics is known for a larger number of positions than with a classical PV diagram. 
\end{itemize}
All the PV diagrams shown in this paper are plotted on a log intensity scale.

We have also derived maps of the continuum, H$\alpha$ emission, and 
velocity field. The analysis of the spectroscopic cubes has shown that in relevant regions (i.e. HII regions of tidal tails), the $3 \sigma$ width of the H$\alpha$ emission line is typically one half of the observed free spectral range. If the width of the emission line was negligible compared to the spectral range, an unbiased estimator of the continuum would be the median of the channel flux. As typically half of the 24 observed channels are found to correspond to the continuum, and the other half to the emission line, we chose to estimate the continuum as the median of the 12 lowest channels. After subtraction of the continuum, the emission line was fitted with a Gaussian, which enabled us to derive the maps of the H$\alpha$ monochromatic emission and of the velocity field.

\begin{figure}[!h]
\centering
\includegraphics[width=8cm]{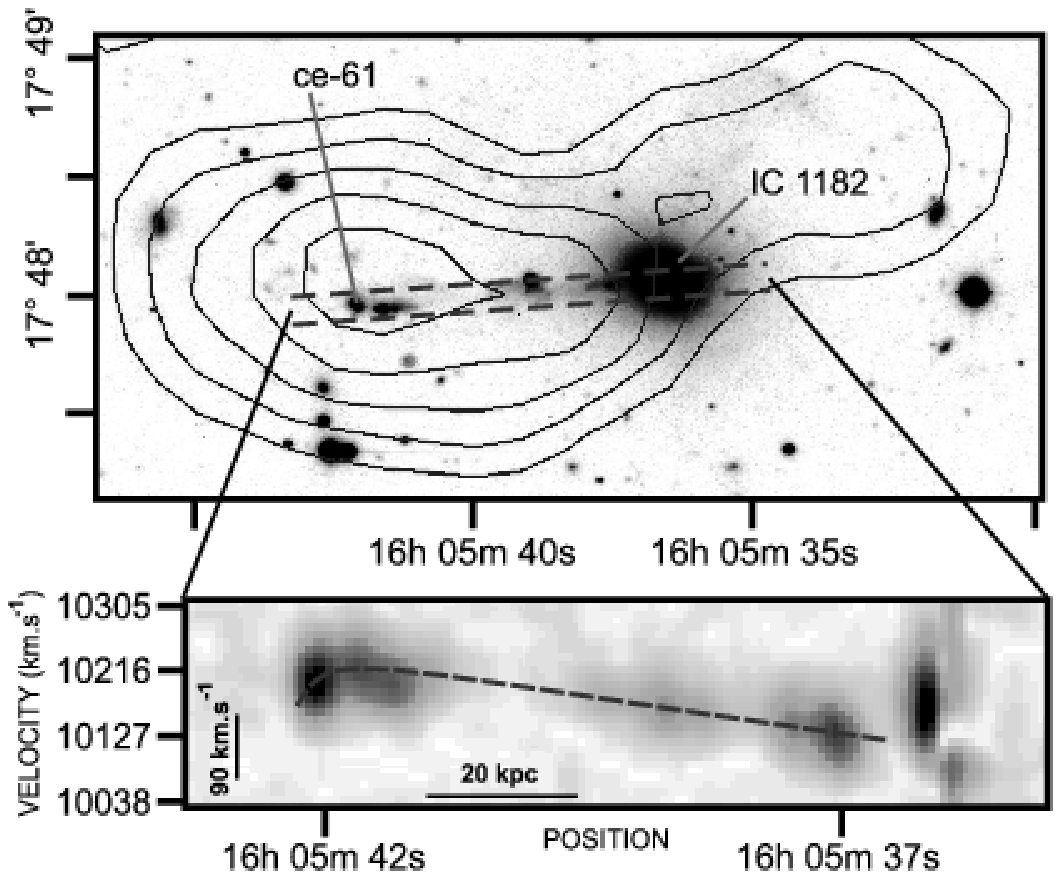}
\caption{Large-scale kinematics of the eastern tidal tail of IC~1182. Top: Optical R-band map (Iglesias-P\'aramo et al. 
2003) and HI contours (Dickey 1997). The TDG candidate ce-61 is identified. The dashed lines delineate the band along 
which the PV diagram has been derived; north is to the top. Bottom: PV diagram of the tidal tail (20 kpc=28''). The dashed line is an eye fit of the H$\alpha$ emission distribution along the tail. 
%symbolizes the interpretation in term of a continuous diagram, while the H$\alpha$ emission is not continuous. 
In addition to this large-scale kinematical signature, each HII region may 
present an inner velocity gradient.}
\label{PV1182}
\end{figure}

\begin{figure}[!h]
\centering
\includegraphics[width=8cm]{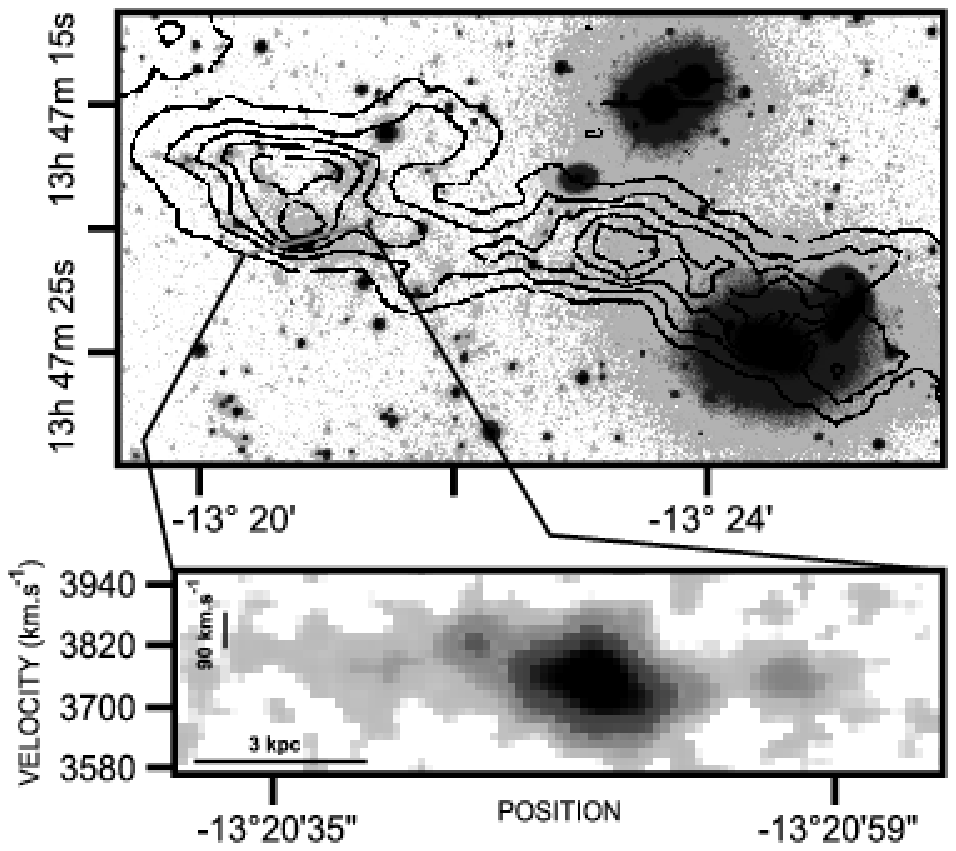}
\caption{Small-scale kinematics of a TDG candidate in the northern part of NGC~5291. Top: optical V-band map (Duc \& Mirabel 1998) and HI contours (Malphrus et al. 1997) ; north is to the left. The dashed line indicates the slit used for the
 PV diagram. Bottom: PV diagram of the TDG candidate (3 kpc=10''), showing an inner velocity gradient 
as large as 100 km~s$^{-1}$ over 2.4~kpc.}
\label{PV5291}
\end{figure}

%\subsection{Results}

In Appendix~\ref{appen} we present a brief description of each observed system, the H$\alpha$ map we have obtained, and some velocity fields.
Two kind of kinematical signatures are observed in our spectroscopic data:
\begin{itemize}
\item large-scale velocity gradients (see example in Fig.~\ref{PV1182}), mainly resulting from streaming motions and the projection of velocities along the line-of-sight
\item inner velocity gradients (see example in Fig.~\ref{PV5291}), possibly related to the rotation of decoupled structures inside tails
%, and other small-scale kinematical structures as dedoubling or uncoupling.
\end{itemize}

The large-scale gradients are analyzed in Sect.~3. We study small-scale kinematical structures in Sect.~4. We have assumed $H_0=70$~km~s$^{-1}$~Mpc$^{-1}$ in these sections.

%%%%%%%%%%%%%%%%%%%%%%%%%%%%%%%%%%%%%%%%%%%%%%%%%%%%%%%%%%%%%%%%%%%%%%%%%%%%%%

\section{Large-scale kinematics: constraints on the orientation of tidal tails}

\subsection{Numerical simulations}
We analyze the large-scale velocity gradients observed along tidal tails  with the help of a large set of numerical simulations of interacting galaxies. Assuming that the large--scale motions are mostly governed by the kinematical response of the gas to tidal interactions, we chose to study the gas kinematics using a restricted three-body code. We checked this hypothesis using a full N-body code.

\subsubsection{Code description}

\begin{figure}[!h]
\centering
\includegraphics[width=8cm]{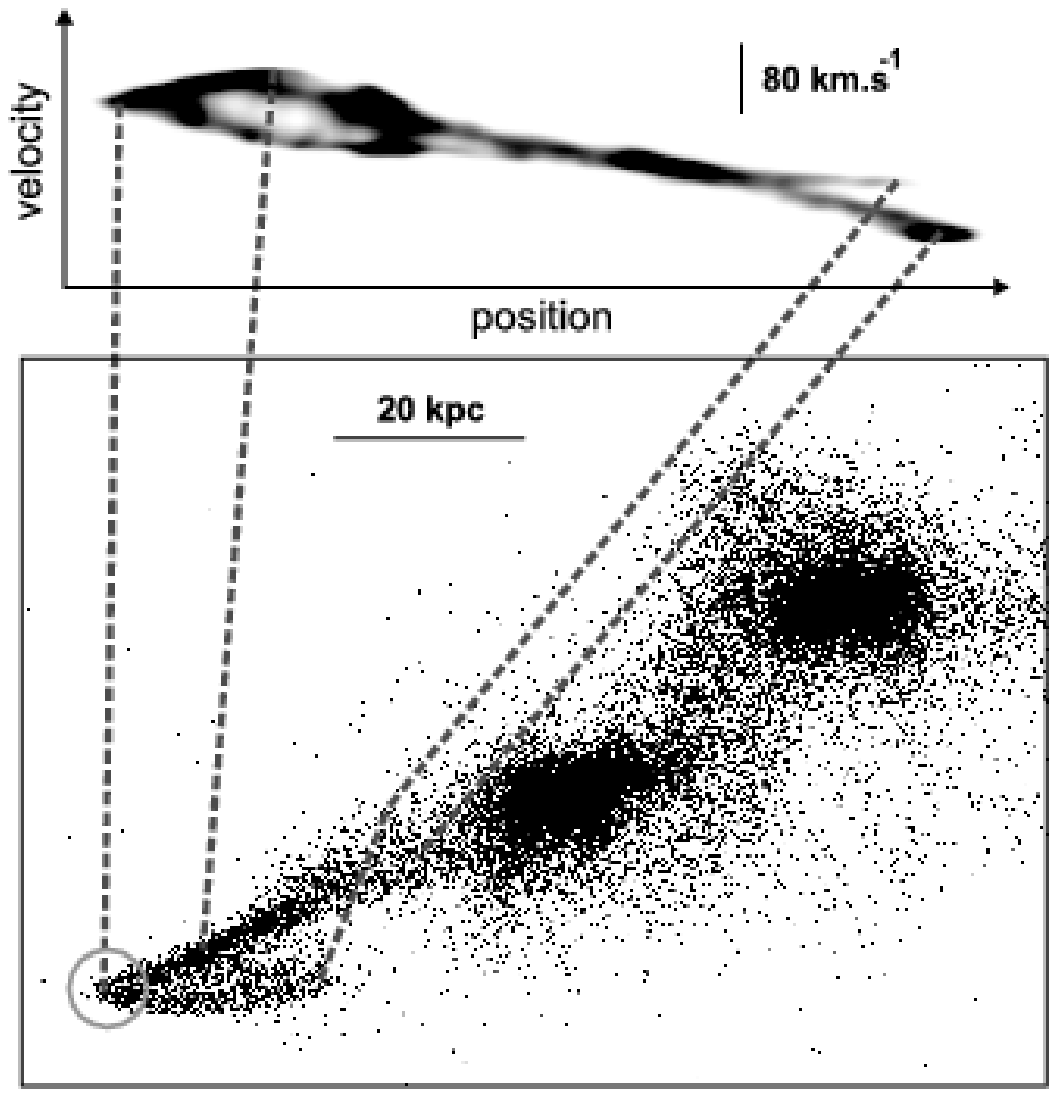}
\caption{Example of a restricted three-body simulation of two interacting galaxies, with a mass ratio of 1:1 (impact parameter 100 kpc, relative velocity 150 km~s$^{-1}$, inclination 30 degrees). In the plot of the projected
mass distribution (bottom), the tidal tail is seen rather close to edge-on (inlination of 65 degrees) whereas the simulated PV diagram (top) corresponds to 
the tail which would be observed exactly edge-on. 
A projection effect exists when part of the tail is aligned with the line-of-sight (encircled region). The velocity gradient is first positive from the parent galaxy to the outer parts, and then becomes negative {\it before} the region affected by the
 projection problem. This early change in the sign of the velocity gradient is the clear signature of a projection
effect at the apparent tip of the tail. Further away, the observed (projected) velocity keeps on decreasing,
drawing a loop in the PV diagram.}
\label{fig4}
\end{figure}

\begin{figure}[!h]
\centering
\includegraphics[width=8cm]{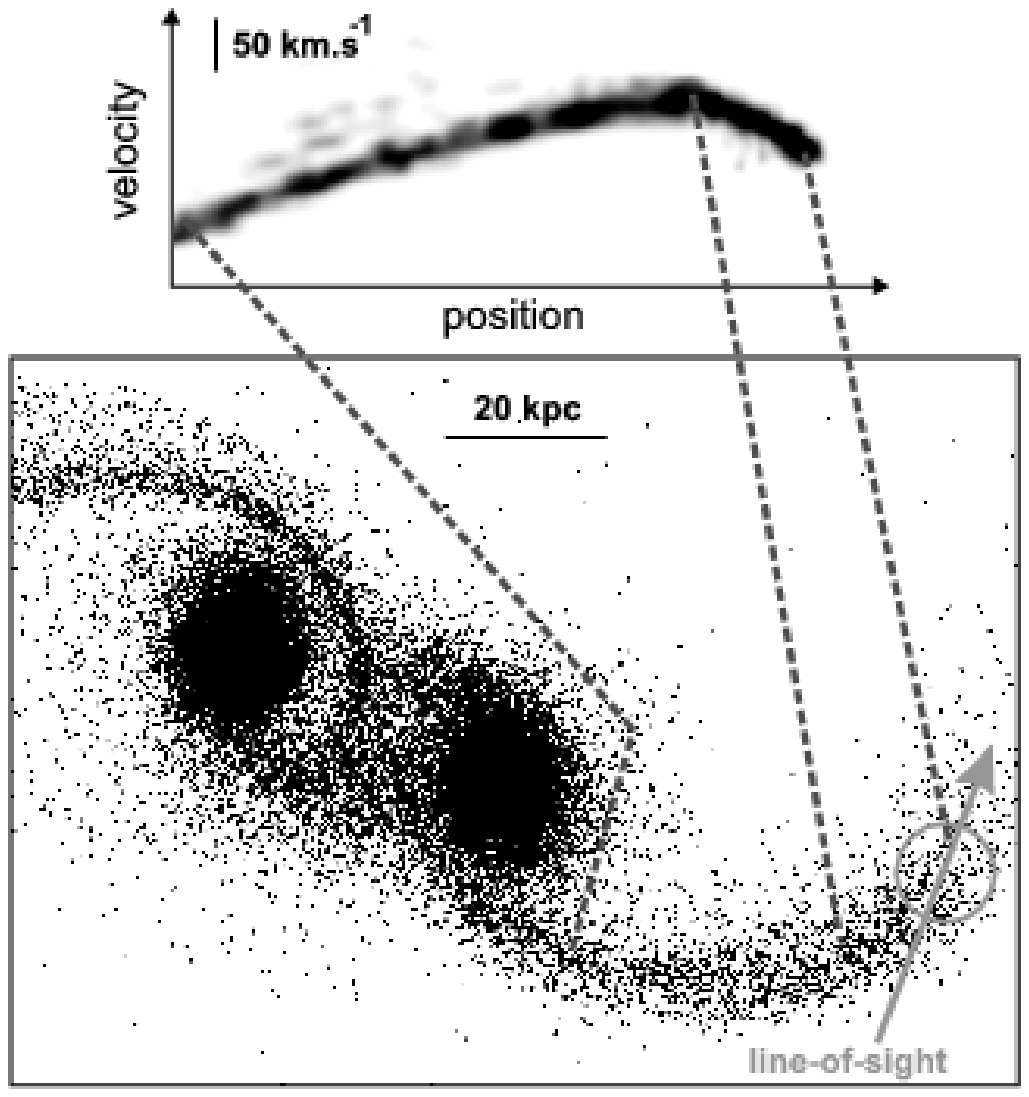}
\caption{Example of a full N-body simulation of two interacting galaxies, with a mass ratio of 1:1 (impact parameter 100 kpc, relative velocity 200 km~s$^{-1}$, inclination 0 degrees).
In the plot of the mass distribution (bottom), the tidal tail is seen 
face-on 
whereas the simulated PV diagram (top) corresponds to the tail which would 
be observed exactly edge-on. 
 A part of the tail is aligned with the line-of-sight (indicated with the arrow), so that a projection effect exists.
In the configuration of this run, the velocity gradient is first positive and then becomes negative before the 
extremity of the tail. }
\label{fig5}
\end{figure}

In our restricted three-body code, two galaxies are tidally interacting on 
a prograde orbit. Stars and dark matter, that dominate the large-scale 
potential, are considered as rigid potential for each galaxy. For a galaxy 
of mass $2\times 10^{11}$ M$_{\sun}$, we chose a rising rotation curve in 
the 500 central parsecs, then a flat rotation curve up to 150 kpc with a circular velocity of 150 km~s$^{-1}$, and a keplerian rotation curve beyong this radius. The gas is initially distributed in a thin Toomre disk, with a scale-length of 15 kpc and a truncation radius of 40 kpc and its initial velocity dispersion is 10 km~s$^{-1}$. For other galactic masses, the lengths are scaled by the square root of the mass. The dissipative dynamics of gas is implemented through the sticky particles-code of Bournaud \& Combes (2002). We have run a set of 54 simulations with various combinations of initial parameters:
\begin{itemize}
\item mass ratios of 1:1, 1:2, 2:1, 1:3 and 3:1
\item impact parameters 50 kpc, 100 kpc, and 150 kpc
\item relative velocities 150 km~s$^{-1}$ and 200 km~s$^{-1}$
\item inclination of the orbital plane of the disturbing galaxy with respect to the analyzed galaxy of 0, 30, and 60 degrees.
\end{itemize}

The main limitations of this model are the absence of self-gravity inside tails, and the modeling of galactic potentials by rigid potentials. Thus, we also simulate tidal tails using the FFT N-body code of Bournaud \& Combes (2003), that models the dynamics of stars, gas, and dark matter, star formation and stellar mass-loss. As for the large-scale kinematics of tidal tails, the N-body code and the restricted three body code give similar results (see Figs.~3 and 4): streaming motions dominate the large scale velocity field. The restricted three body code is thus adapted to the problem studied here.

\subsubsection{Results of numerical models}\label{simul}

We have simulated major galactic encounters, then reproduced the 
observation of tidal tails that are seen edge-on: in this observational 
context, the actual velocity is unknown. We only measure the projection of 
it along the line-of-sight. In Fig.~\ref{fig4}, we show the result of a 
restricted three-body simulation, and give the PV diagram expected when a 
part of the tail is aligned with the line-of-sight, i.e. when an observed 
condensation at the tip of the tail may in fact be the result of a 
projection effect. We show in Fig.~\ref{fig5} a similar result with  a 
N-body simulation. In Fig.~5, we illustrate how a projection effect may result in an apparent accumulation of matter.

The result of all our simulations, is that {\it the PV diagram of a tidal 
tail for which a projection effect exists shows a change in the sign of 
the velocity gradient}. If the initial velocity gradient of the tail is 
positive (resp. negative), the projected velocity reaches a maximum 
(minimum) value before the end of the projected tail, then the sign of the 
velocity gradient changes, and the velocity at the tip of the projected 
tail is smaller (larger) than the maximum (minimum) velocity. In 
Fig.~\ref{fig4}, we have identified several points in the tidal tail and 
reported them on the PV diagram, which clearly shows that the maximum of 
the projected velocity is reached before the apparent tip of the tail. 
When the tidal tail extends further than the location of the projection 
effect, the PV diagram has a loop shape (see the case of Fig.~\ref{fig4}). 
On the contrary, a velocity gradient that does not change its sign along a 
tidal tail would prove that no projection effect is present in this tail.

\begin{figure}
\centering
\includegraphics[width=8cm]{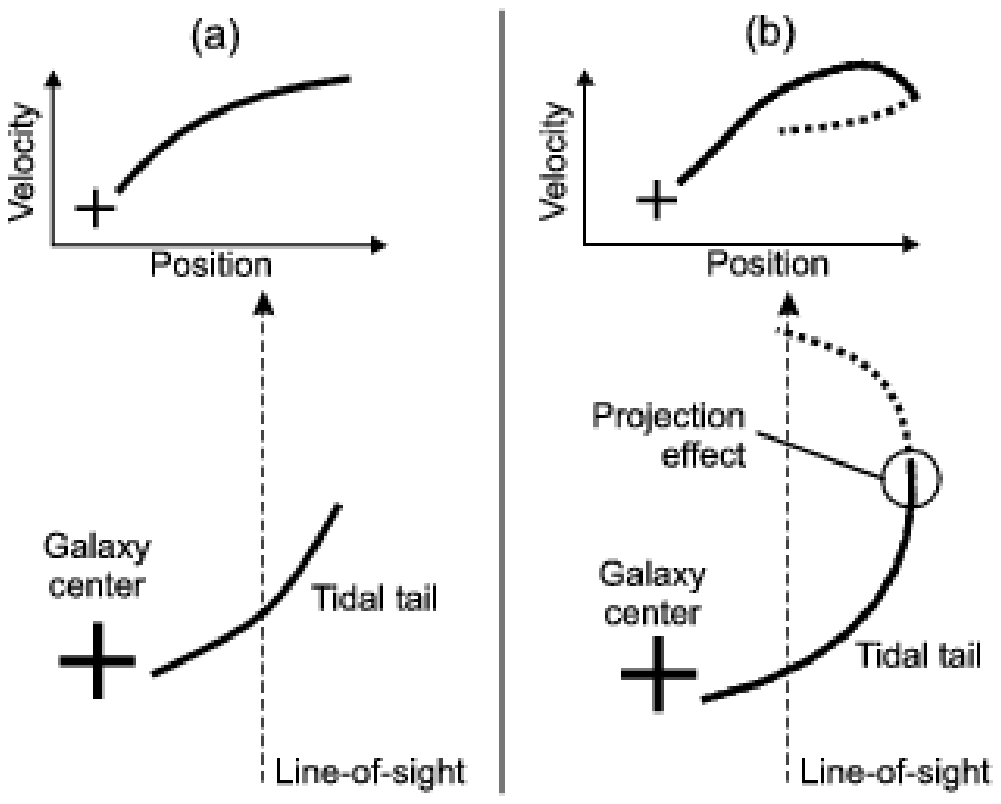}
\caption{PV diagrams expected for tidal tails that are seen edge-on, according to the results of our numerical
 simulations. Case (a): no part of the tail is aligned with the line-of-sight and an observed mass concentration cannot 
result from a projection effect. The sign of the projected velocity gradient does not change. Case (b): a projection effect exists 
and could be responsible for the observation of a mass concentration. The sign of the velocity gradient changes before the apparent extremity of the tail. A loop-shaped diagram is obtained when the tidal tails extends much further than 
the projection effect (dashed lines). The whole loop may however be difficult to detect since the most external parts of
tidal tails are likely to be faint.}
\label{schema2}
\end{figure}

For the 54 restricted three-body simulations, we derived the PV diagram of tails
 observed with a line-of-sight that is aligned with a part of them (i.e. a projection effect exists).
We obtained the following statistical results on the projected velocities:
\begin{itemize}
\item the total velocity extension along the tidal tail starting from the disk is 190 $\pm$ 70 km~s$^{-1}$
\item the extremum (maximum or minimum value depending of the orientation) velocity is reached 
at 80 $\pm$ 10 \% of the projected tail length.
\item the mean velocity at the tip of the projected tail is 50 $\pm$ 30 km~s$^{-1}$ smaller/larger than the 
extremum value.
\end{itemize}
According to these values, the change in the sign of the large-scale velocity gradient related to the projection effect should be detected in our observations. We have also varied the gas mass and radial distribution in the initial disk: this does not influence the kinematical signature of projection effects. In the following we apply these results, summarized in Fig.~\ref{schema2}, to the systems for which we have obtained Fabry-P\'erot data and for which tidal tails are observed edge-on or nearly edge-on. We did not attempt though to fit the morphology and kinematics of each system with ad hoc simulations.

\subsection{Application to observed systems}
A brief description of all observed systems is given in Appendix~\ref{appen}. We only discuss here the objects for which the ionized gas could reliably trace the global kinematics of the tidal tails, i.e. enough HII regions are detected. Existing HI data were also used.

\paragraph{IC~1182:}
The presence of a TDG candidate, ce-61, is reported at the extremity of the eastern tail of IC~1182 (Braine et al. 2001; Iglesias-P\'aramo et al. 2003). The kinematical data for this tail are shown in Fig.~\ref{PV1182}. The projected velocity first rises away from the IC~1182 center with an amplitude of 130 km~s$^{-1}$, then decreases by 40--50 km~s$^{-1}$ near its extremity. This kinematical signature corresponds to case (b) in Fig.~\ref{schema2}, indicating that the end of the tail may be aligned with the line-of-sight. The candidate TDG observed is this region then seems to result from a projection effect, even if a genuine object could exist in addition to the apparent condensation.

\begin{figure}
\centering
\includegraphics[width=8cm]{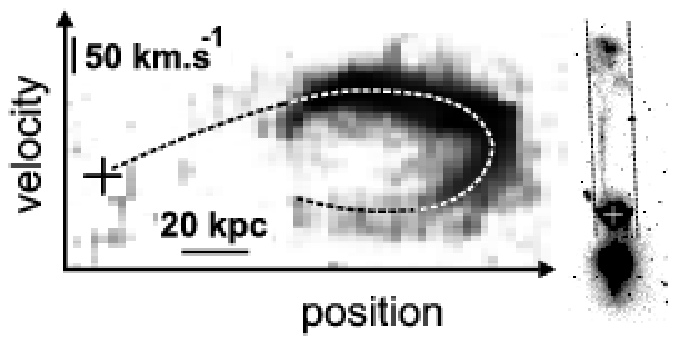}
\caption{PV diagram of the northern tidal tail of Arp~105, from HI data of 
Duc et al. (1997) (20 kpc=33''). We find here the kinematical signature 
expected for a tidal tail with a projection effect (case (b) in 
Fig.~\ref{schema2}). The band along which the PV diagram has been 
integrated is shown in the optical map. The crosses on the diagram and the 
optical map identify the position of the center of the parent spiral 
galaxy.} \label{Arp105N}
\end{figure}

\paragraph{Arp~105:}
Two tidal tails and TDG candidates were observed in that system. 
Studying the orientation and internal structure of the northern tail, Duc et al. (1997) concluded that a projection effect may explain part of its properties. Indeed, the PV diagram of the HI component, integrated along the full width of the tail (see Fig.~\ref{Arp105N}), shows the kinematical signature expected from simulations: a loop feature similar to case (b) in Fig.~\ref{schema2} with characteristic amplitudes consistent with the values indicated in Sect.~3.1. If a real object is present there in addition to the projection effect, its mass has been over-estimated.

\begin{figure}
\centering
\includegraphics[width=8cm]{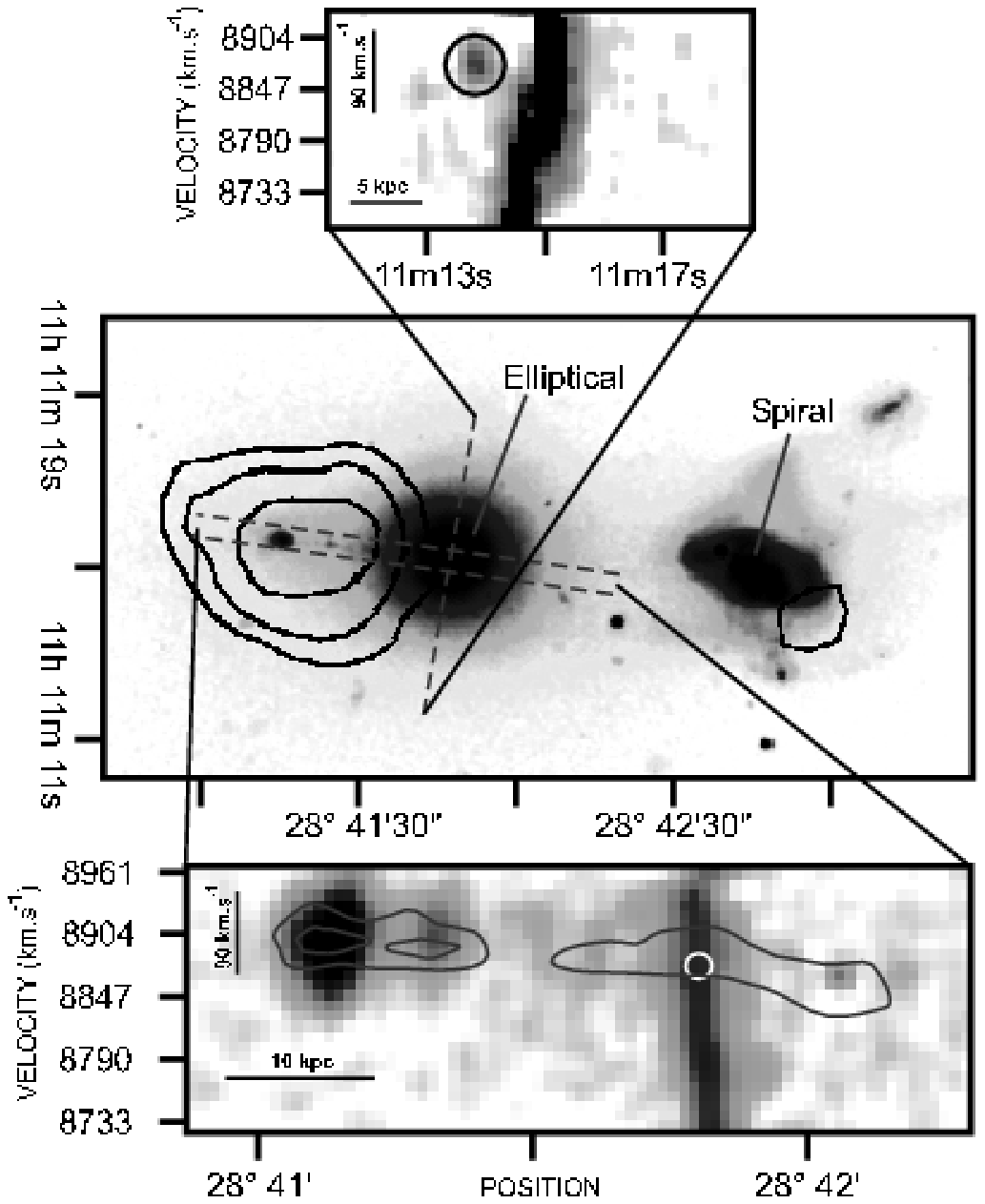}
\caption{Kinematics of the southern tail of Arp~105 (NGC 3561, The Guitar). Center: Optical V-band map (Duc \& Mirabel 1994) and HI contours (Duc et al. 1997). North is to the right. The TDG candidate is the most luminous feature on the left of the elliptical galaxy. 
Top: H$\alpha$ PV diagram, and identification of a HII region hidden by 
the elliptical but kinematically decoupled from it (circled region); the 
velocity of this region coincides with the kinematics of the tidal tail issued from the spiral, so that this region is likely to belong to the tail passing in front of the elliptical. Bottom: Large-scale H$\alpha$ PV diagram of the tidal tail (10 kpc=16''). The circle indicates the mass center of the HII region identified in front of the elliptical galaxy (top image). The grey contours corresponds to the kinematics of the HI component, according to the B--array VLA observations of Brinks et al. (2004). The large-scale kinematics of this tail indicates that the TDG candidate does not result of a projection effect, but is a genuine accumulation of matter.}
\label{PV105}
\end{figure}

The PV diagram of the southern tail of Arp~105 is shown in Fig.~\ref{PV105}. 
This tail emanates from the spiral galaxy, passes in front of the elliptical galaxy and 
ends in a compact blue object. The velocity gradient along the tail, traced by several HII regions, is positive from the spiral galaxy outwards and never changes its sign. High resolution HI data (Brinks et al. 2004, in prep.)  are compatible with the kinematics of the HII regions. 
%the maximal projected velocity is observed at the extremity of the tail.
 The HII and HI kinematical data correspond to case (a) in Fig.~\ref{schema2} where no projection effect is present. Thus, the massive condensation of HI, CO and stars, observed at the extremity of this tidal tail (Duc \& Mirabel 1994; Duc et al. 1997; Braine et al. 2001) cannot be caused by a projection effect, but is on the contrary a genuine object.

This conclusion assumes that the HI and HII emission observed in the southern parts of Arp~105 are related to a tidal tail emanating from the spiral. Other interpretations might be proposed, such as a jet emitted by the elliptical galaxy (Arp 1966; Arp \& Madore 1967), or a polar ring forming around the elliptical galaxy, following a process detailed in Bournaud \& Combes (2003). Duc et al. (1997) presented several lines of evidence that this tail is really a tidal structure. They also ruled out the hypothesis that the TDG candidate could be a pre-existing object that has attracted the tidal tail to it.

\paragraph{Arp~242:}

\begin{figure}
\centering
\includegraphics[width=8cm]{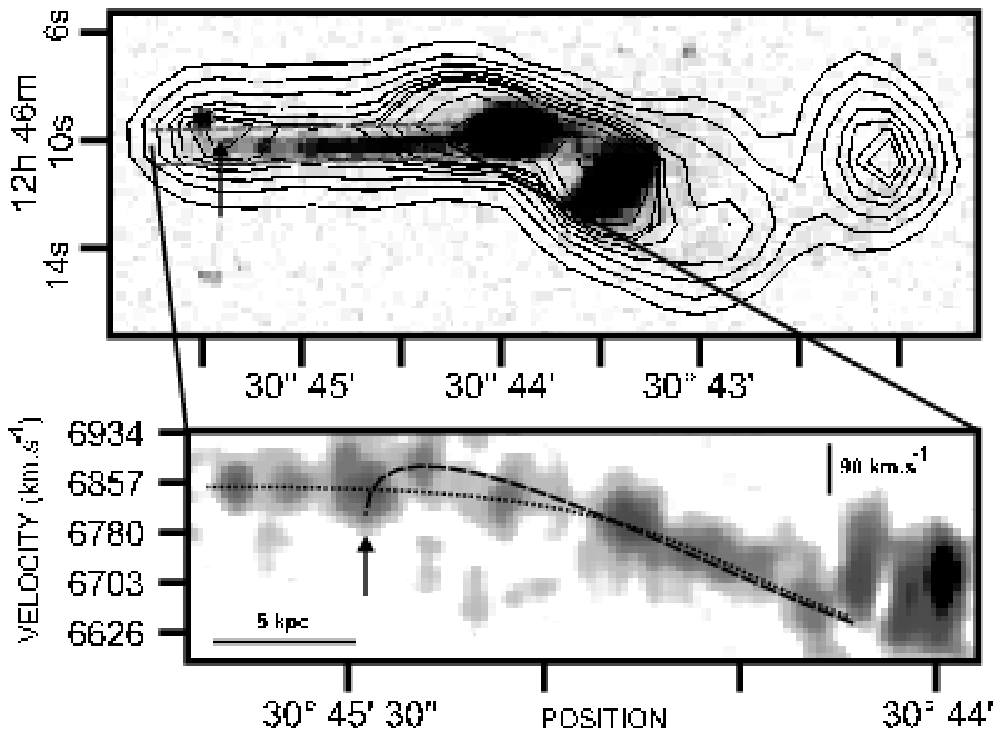}
\caption{Kinematics of the northern tail of Arp~242. Top: DSS blue image and HI contours (Hibbard \& van Gorkom 1996). North is to the left; the arrow identifies the TDG candidate. -- Bottom: H$\alpha$ PV diagram of the northern tail (5 kpc=11''). The arrow indicates the H$\alpha$ peak of the TDG candidate. One could attempt to interpret the kinematics of the tail with a change in the sign of the velocity gradient and a projection effect associated with the TDG candidate (long-dashed line), but the last 5 kpc of the tail are not fitted, which rules out this interpretation. On the contrary, a more robust interpretation of the kinematics is obtained without any change in the sign of the velocity gradient, and only one HII region that is not perfectly fitted (short-dashed line), suggesting that the mass concentration in this tail is not the result of a projection effect.}
\label{PV242}
\end{figure}

The northern tidal tail of the interacting system Arp 242 (NGC 4676), also known as ``The Mice'',
is seen almost perfectly edge--on. A massive HI condensation is found at its tip (Hibbard \& van Gorkom 1996).
We show in Fig.~\ref{PV242} the H$\alpha$ PV diagram of this tail. The large-scale velocity gradient along the tail has an amplitude of 200 km~s$^{-1}$. As shown in this figure, an interpretation with a change in the sign of the velocity gradient before the candidate TDG would not fit the kinematics of the 5 last kpc of the tail, while an interpretation without any change in the sign of the velocity gradient is in good agreement with the observations; only one faint HII region, located 1 kpc before the center of the TDG candidate, is not exactly fitted in this interpretation. It is thus unlikely that the TDG candidate observed in the northern tail of Arp~242 is the result of a projection effect.
 
\paragraph{NGC~7252:}
The late-state merger NGC~7252 contains a massive TDG candidate in its north-western tidal tail. The
latter is located at a distance of 80 kpc from the parent galaxy, at the extremity of the stellar tail 
and just before a significant decrease in the observed HI column density 
(Hibbard et al. 1994)\footnote{A faint HI tidal extends much further than the optical counterpart}. The linear structure of the stellar tail
raises again the question of projection effects. Hibbard et al. (1994) 
obtained HI spectroscopic observations of this system. They present a PV 
diagram between the origin of the tail and
 the TDG candidate similar to those we made for other systems (see their Fig.~4).
The absence of a change in the sign of the large-scale velocity gradient between the parent galaxy and the 
TDG candidate indicates that projection effects cannot play a major role at the location of the TDG candidate. More precisely, a change in the sign of the velocity gradient is actually seen in the observations of Hibbard et al. (1994), but it occurs at radii larger than the position of the TDGs. Thus, projection effects may exist in the faint HI outer tail, but not near the TDG candidate.

In the eastern tail of NGC~7252, a massive TDG candidate, very similar to that present in the north-western tail, is also observed (Hibbard et al. 1994). According to the HI kinematical data of Hibbard et al. (1994), the velocity gradient is nearly constant between the parent galaxy and this TDG candidate, so that this object is likely a real condensation of matter. As in the north-western tail, projection effects associated with a change in the sign of the velocity gradient probably occur in the faint outer tail, further than the TDG location, but cannot be responsible for the observed accumulation of matter.

\subsection{Conclusion: projection effects versus genuine objects}
The study of the large-scale velocity gradient of tidal tails and the
 comparison with numerical models has enabled us to get information
about the three-dimensional geometry of tidal tails. The kinematics of the eastern tidal tail of IC~1182 is fully compatible with a projection effect along the line-of-sight at the extremity of the tail. The TDG candidate ce-61 could then be the result of a projection effect, but the presence of a genuine concentration of matter cannot be ruled out. A projection effect is also present in the northern tail of Arp~105: this tail may still contain a genuine accumulation of matter, but its mass would have been largely overestimated. On the other hand, in the cases of Arp~105 South, Arp~242 North, and NGC~7252 (East and North-West), we have shown that the accumulations of matter of typical masses of  $10^9$ M$_{\sun}$, observed far from the parent galaxies, cannot be associated with projection effects.

The case of NGC~5291 has not been discussed in this Section. Indeed the HI structures in this system are likely not to have the same origin as usual tidal tails: the large HI ring observed in this system was probably formed during a high-velocity head-on galaxy collision (Appleton \& Struck-Marcell 1996), as for the Cartwheel galaxy (e.g., Horellou \& Combes 2001). The massive accumulations of matter detected by Duc \& Mirabel (1998) in the ring are nevertheless  relevant to our study of the genesis of dwarf galaxies in the large-scale debris of  a galactic encounter. In that system, the massive condensations of matter cannot result from projection effects for the simple reason that they are located far from the apparent extremities of the ring major axis where projection effects would be located.

The other criterion put forward to disentangle real objects and the results of projection effects is the detection of the millimetric CO line, if formed locally out of condensed HI (Braine et al. 2000, 2001). It turns out that this line has been detected towards all objects for which we claim that projection effects do not play a major role (Arp 105--South, Arp 142--North, NGC 7252, NGC 5291). On the other hand, no molecular gas was detected towards the eastern tail of IC 1182 and the northern tidal tail of Arp 105 where, based on our kinematical data alone, we found evidence for projection effects. We thus find a trend between the detection of CO and the absence of projection effects. However, based only on our restricted sample, it is premature to infer that this is a real correlation.

Projection effects occur in interacting systems, and some observed accumulations of matter are only the result of them. We have however shown that genuine massive accumulations of matter exist too. 
They could thus be the progenitors of Tidal Dwarf Galaxies. This answers the first concern raised in the introduction. 
We are now to examine the small-scale kinematics of these TDG candidates, in order to study whether they are self-gravitating, kinematically consistent structures or only transient accumulations of matter, and whether they contain dark matter.

%%%%%%%%%%%%%%%%%%%%%%%%%%%%%%%%%%%%%%%%%%%%%%%%%%%%%%%%%%%%%%%%%%%%%%%%%%%%%%

\section{Small scale kinematics: effects of the self-gravity of tidal tails}
\subsection{Inner kinematical structures}

%In the previous section, we have studied the velocity gradient of tidal tails at scales of a few kpcs. 
After having examined the velocity gradient of tidal tails at scales of tens of kpc, we now examine the inner kinematics of the individual giant HII regions at scales of a few hundreds of pc. The main objectives are to determine if the condensations  are kinematically decoupled from the rest of the tail, estimate their dynamical mass and study their dark matter content. To study their inner dynamics, we first subtracted the large scale velocity gradients associated with streaming motions.

In the southern TDG candidate of Arp~105 -- a genuine object, as argued in Sect.~3 -- a velocity gradient seems to be detected but not fully resolved (see Fig.~\ref{anx105}). It could be a velocity gradient of up to 100 km~s$^{-1}$ over 2 to 3 kpc (in addition to the large-scale gradient of the tidal tail, shown in Fig.~\ref{PV105}). However, as seen on this figure, the spatial and spectral resolution are too low to derive the exact value of this gradient. 

%The confirmation of the presence of the inner velocity gradient that seems to be present but not fully resolved in our %observations, would definitively prove that this object is an actual tidal dwarf galaxy, i.e. a real condensation of matter %that is self-gravitating.

In IC~1182, a giant HII region located close to the TDG candidate seems to exhibit a similar small-sale velocity gradient, in addition to the tail large-scale gradient, with a maximal value of 70 km~s$^{-1}$ accross 2~kpc (see Fig.~\ref{anx1182}) but is not fully revolved. As noted in Sect.~3.2, in this system projection problems may hamper the interpretation. 

In the case of NGC~5291, the evidence for the presence of inner velocity gradients is much more robust. In the southern regions, a velocity gradient of 50 km~s$^{-1}$ over 2kpc is detected (see Fig.~\ref{anx5291S}), but it does not coincide with in the brightest HII region. In the northern regions, the main HII region -- the most luminous TDG candidate in the system 
(Duc \& Mirabel 1998) -- shows a velocity gradient as large as 100 km~s$^{-1}$ over 2.4 kpc (see Fig.~\ref{PV5291}). This genuine 
accumulation of matter (see Sect.~3.3) appears to be self-gravitating and 
kinematically decoupled from the neighboring 
regions.
% thus to be actually a dwarf galaxy formed in the remnant of a galactic encounter. 
The implications of this gradient on the dynamics of the TDG and its dark matter content are discussed in Sect.~\ref{gradient}. 

In general, we confirm with our Fabry--P\'erot observations the existence of the inner velocity gradients
that had previously been found at the same positions in long-slit spectroscopic observations 
(see Duc et al. 1997 for Arp~105, and 
Duc \& Mirabel 1998 for NGC~5291). The values of the velocity gradients derived with the FP instrument,
although still uncertain, are however smaller than the values initially reported: 50--100 km~s$^{-1}$
instead of 100--150  km~s$^{-1}$. The discrepancy most probably comes from the different spectral
resolutions.

\subsection{Dark matter in tidal dwarfs: the case of NGC~5291}\label{gradient}
The TDG candidate North of NGC~5291 is the only object in our sample for which the
inner kinematics is sufficiently spatially resolved to measure with reasonable precision the 
 velocity gradient and estimate from it a dynamical mass. 
We obtained a value of 100 $\pm$ 20 km~s$^{-1}$ (6 channels) over 2.4 $\pm0.4$ kpc (20 pixels) (see Figs.~\ref{PV5291} and \ref{rot_curve}). This gradient is consistent with the one initially derived by Duc \& Mirabel (1998) from long-slit
spectroscopy (however over a larger region and a slightly different orientation).

\begin{figure}
\centering
\includegraphics[width=8cm]{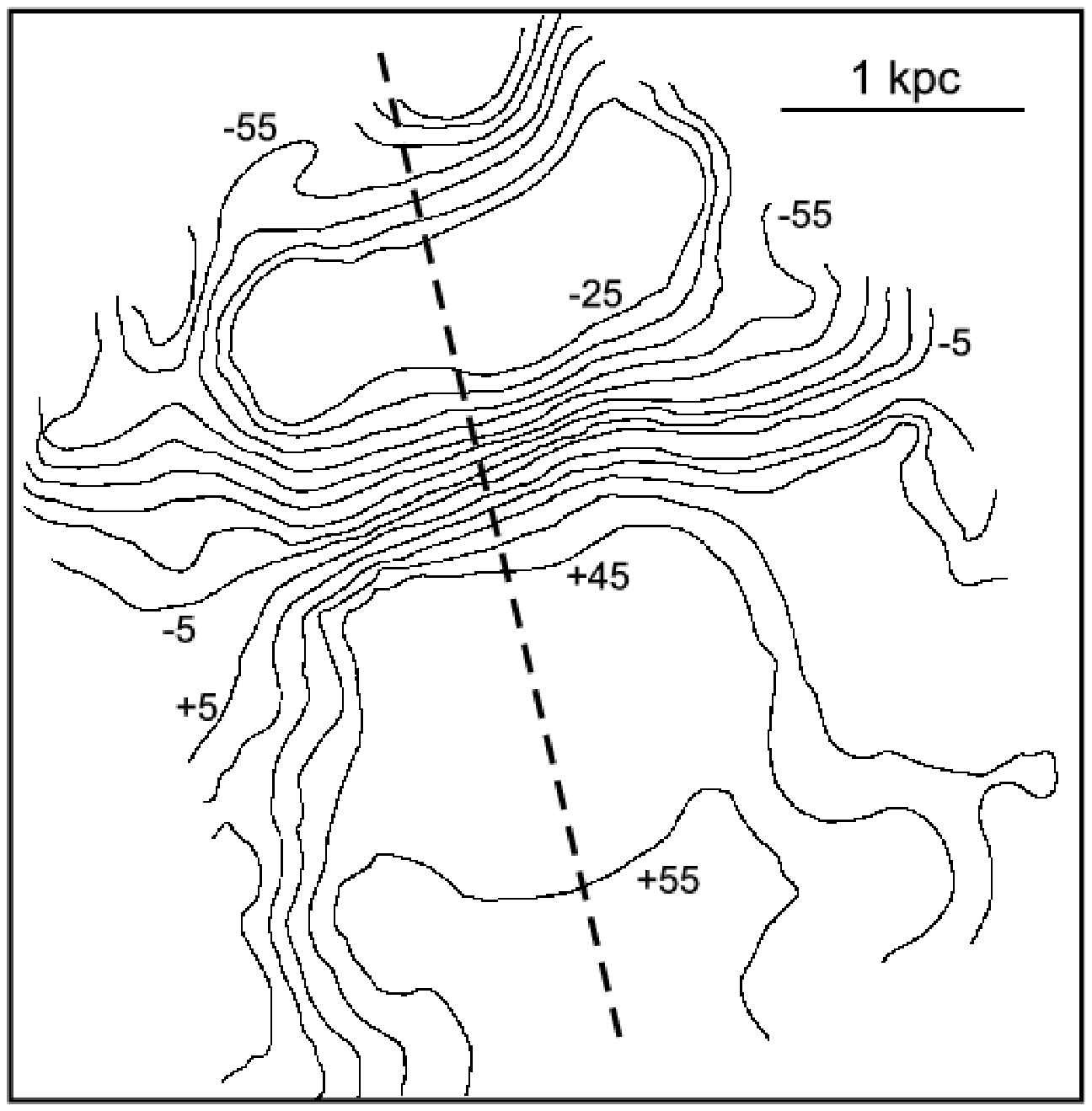}
\caption{Isovelocities diagram of NGC~5291 northern TDG (1 kpc=3''). The classical spider shape, corresponding to the rotation of the system, can be identified, but is largely disturbed which probes that supplementary motions or distortions play a role.}
\label{isoV}
\end{figure}

\begin{figure}
\centering
\includegraphics[width=8cm]{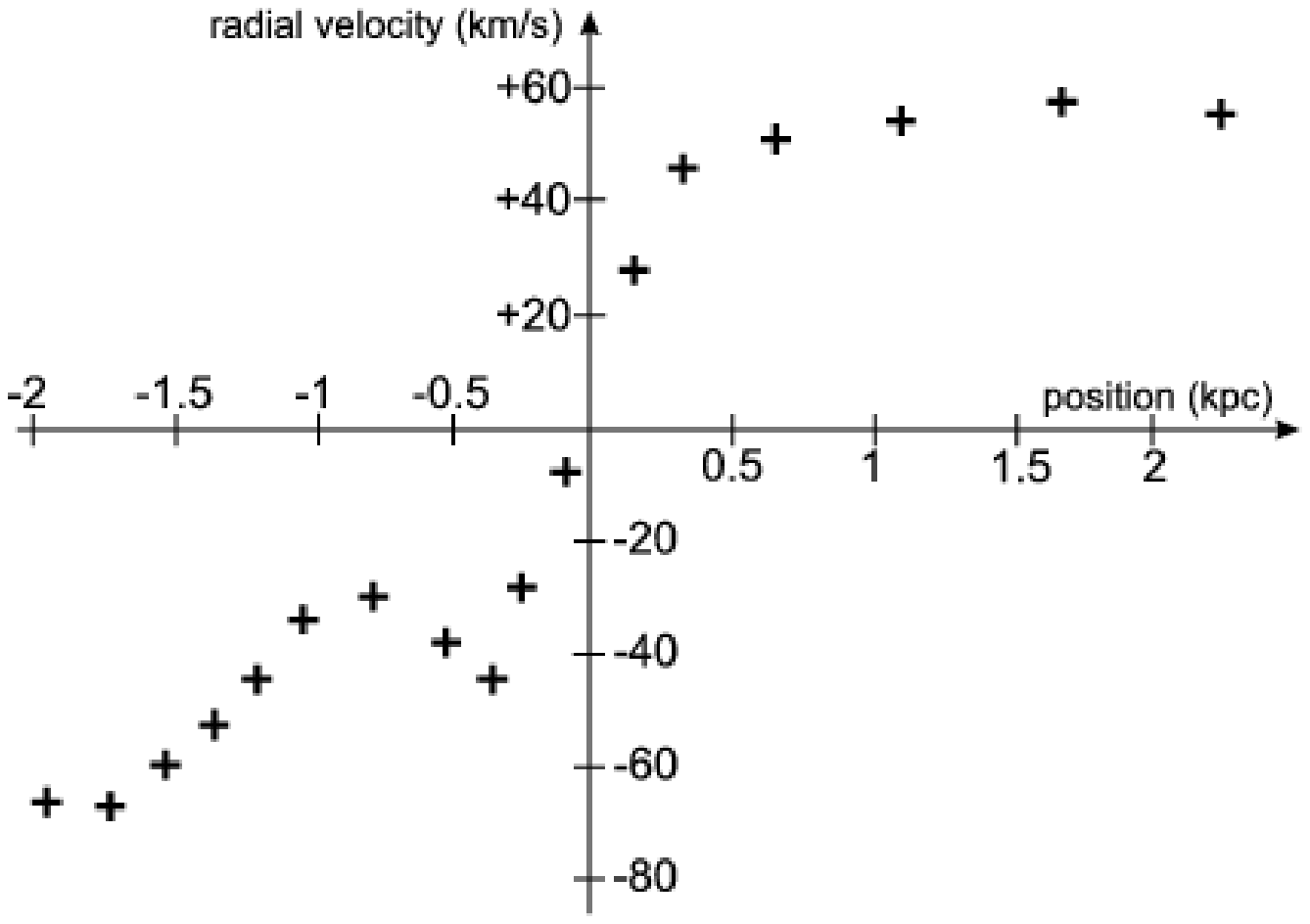}
\caption{Projected velocity of the northern TDG of NGC~5291, as a function of the position along the axis shown in Fig.~\ref{isoV} (1 kpc=3''). The center of rotation is here defined as the mass center of the H$\alpha$ emission, which is very
 close to the H$\alpha$ emission peak. The right part of this curve, corresponding to the southern part of the TDG, is similar to the classical rotation curves of disk galaxies while the left (northern) part seems very disturbed.}
\label{rot_curve}
\end{figure}

The velocity field, shown in Fig.~\ref{isoV} (see also 
Fig.~\ref{anx5291N}), globally presents the classical spider-shape of 
rotation. Dynamical disturbances and asymmetries are however observed. The 
presence of a warp resulting in a non-constant projection factor could 
account for the perturbation. Indeed, this system is believed to have 
formed very recently, or to be still forming through gas accretion, thus 
is very likely to be warped. In Fig.~\ref{rot_curve}, we show the 
projected velocity as a function of the position along the axis shown 
in Fig.~9 crossing the center of this giant HII region. In the southern 
part, we 
get a flat curve, rather similar to rotation curves of spiral galaxies, while the northern part shows a globally rising but disturbed curve. 

We can derive a dynamical mass, assuming in a first approach that the 
observed velocity gradient has a purely rotational origin. The large-scale 
HI structure of NGC~5291 is a ring (Malphrus et al. 1997), inclined by 
about 45 degrees with respect to the plane of the sky. As a first 
approximation, one may assume that the TDG gaseous disk has also the same inclination. The isovelocity diagram (Fig.~\ref{isoV}) is actually consistent with a disk 
%that is seen both far from edge-on and far from face-on, 
and the ellipticity of the outer H$\alpha$ isophotes indicates an inclination of 40 to 50 degrees. 
The deprojected velocity gradient is then 140 km~s$^{-1}$ over a distance 
of 2.4 kpc. The associated dynamical mass is $M_{\mathrm dyn} = R V^2/G = 
1.4 \times 10^9 \mathrm{M}_{\sun}$ for a spherical distribution. We used a 
radius $R=1.2$ kpc and a velocity $V=70$ km~s$^{-1}$ to derive this value. This mass would be smaller for a flattened distribution.
%\begin{equation}
%M_{\mathrm dyn}=\frac{R V^2}{G}
%\end{equation}

The HI mass inside this system, derived from observations of Malphrus et 
al. (1997) is $M_{\mathrm {HI}} \simeq 2.6 \times 10^9 \mathrm{M}_{\sun}$. 
If 
we add the mass of stars, derived from an optical blue image, assuming 
M/L=2, we get the total mass of visible matter \footnote{A precise value 
of M/L cannot be given without studying in detail the stellar population 
of the system; however the mass is dominated by the HI component.}: 
$M_{\mathrm vis} \simeq 2.8 \times 10^9 \mathrm{M}_{\sun}$.

We would then find an unrealistic visible mass larger than the dynamical 
mass. However, one should notice that the kinetic energy has been 
under-estimated. We have only taken into account the component in rotation 
but not at all the velocity dispersion and the possible global contraction 
or expansion of the system. All these non-circular motions also contribute 
to the kinetic energy. The dynamical mass we have derived is then a lower 
limit to the actual dynamical mass. Braine et al. (2001) obtained a much 
higher estimate of the total dynamical mass of the TDG candidate, based on 
the width of the CO line detected towards it. Arguing that the CO has 
formed in situ, they claim that it traces the true potential well of the 
object. They measured a rather uncertain FWHM of 130 km~s$^{-1}$. According to our determination of the rotation velocity, the non-circular motions would then have an equivalent FWHM of 100 km~s$^{-1}$, so that their contribution to the dynamical mass is of the same order as the contribution of rotation. This finally gives a dynamical mass of $M_{\mathrm dyn} \simeq 3\times 10^9 \mathrm{M}_{\sun}$.

It now seems that $M_{\mathrm vis} \simeq M_{\mathrm dyn}$ which would 
indicate that the system does not contain dark matter inside a radius $r 
\simeq 2$ kpc. Yet, the HI mass is likely to be over-estimated, for:
\begin{itemize}
\item the beam of HI observations may include other regions than the TDG.
\item the dynamical mass of the TDG is estimated from H$\alpha$ observations, while in most galaxies the HI disk is more extended than the area of star formation, so that only a part of the HI mass should be compared to the dynamical mass.
\end{itemize}
%We should thus replace the previous equation by: $M_{\mathrm vis} \leq M_{\mathrm dyn}$. 
Two situations are then possible: either $M_{\mathrm vis} = M_{\mathrm 
dyn}$, the system does not contain dark matter inside radius $r \simeq 2$ 
kpc, or $M_{\mathrm vis} < M_{\mathrm dyn}$, the system contains dark matter.

In the second situation, the discrepancy between $M_{\mathrm vis}$ and $M_{\mathrm dyn}$ is expected to be rather small. We test the presence of dark matter inside the stellar (HII) radius of the system, thus the dark-to-visible mass ratio will not be as high as if we were probing larger radii. If the TDG contains dark matter, the over-estimation of the HI mass in our data may easily hide it, for the excess of $M_{\mathrm dyn}$ with respect to $M_{\mathrm vis}$ will only be a few tens of percent. A factor as large as ten between $M_{\mathrm dyn}$ and $M_{\mathrm vis}$ is only expected if the dynamics is studied at very large radii. For instance in spiral galaxies, the dark-to-visible mass ratio inside the stellar disk radius is typically smaller than 1, and reaches the typical value of 10 only at several times the stellar disk radius (e.g., Persic \& Salucci 1990).

To draw conclusions on the presence of dark matter, higher resolution observations are clearly needed. The estimation of the dynamical mass should be refined by deriving the rotation curve of the system (and not only a typical value of the integrated velocity gradient), its velocity dispersion, and its global contraction or expansion. To this aim, we estimate that a spatial resolution three times better ($\sim$ 100 pc) and a spectral resolution two times better ($\sim$ 8 km~s$^{-1}$) are needed. At the same time, the HI distribution should be observed with the best resolution available, in order not to over-estimate the HI mass of the TDG. The morphology of the system should also be studied in detail, to determine its inclination and derive the deprojected gradient. High resolution stellar or CO observations could help to derive the inclination of the TDG. Finally, a robust model should include the influence of the wide HI ring: this massive non-spherical structure, in which the TDG is embedded, contributes to the dynamics of the TDG that may play a role. Indeed, the flattening of the HI distribution in the ring may generate asymmetries in the TDG dynamics, such as eccentric orbits. This could be studied using models of the gravitational collapse of a TDG in a massive ring.

%%%%%%%%%%%%%%%%%%%%%%%%%%%%%%%%%%%%%%%%%%%%%%%%%%%%%%%%%%%%%%%%%%%%%%%%%%%%%%

\section{Conclusions}

Using numerical simulations, and taking advantage of the large field of view and high spatial resolution of the Fabry-P\'erot observations, we have shown that:
\begin{enumerate}
\item The large-scale kinematics of tidal tails can help to disentangle genuine condensations of matter and apparent condensations resulting of projection effects at the extremity of tidal tails.
\item Several accumulations with typical masses of $10^9$ M$_{\sun}$ observed
at the extremity of tidal tails cannot be the result from projection effects
but are real condensations of matter. It is the case for Arp~105--South, Arp~242--North,
NGC~7252, and NGC~5291--North. On the other hand, the kinematics of the eastern tail of IC~1182 and the northern tail of Arp~105 are compatible with projection effects: in these cases, the apparent accumulations of matter as massive as $10^9$ M$_{\sun}$ can result of the over-estimation of the mass of much smaller condensations (Hibbard \& Barnes 2004). It is not surprising that projection effects occur in some systems, but genuine accumulations of matter are proven to exist in several systems. Such condensations as massive as 10$^9$~M$_{\sun}$ are also observed in tidal tails seen more or less face--on, for instance in NGC~7769/7771 or Arp 299 that can be seen in the maps of the HI Rogues Gallery\footnote{available at http://www.nrao.edu/astrores/HIrogues/} (Hibbard et al. 2001).
\item Velocity gradients are detected inside the most luminous tidal HII regions, with amplitudes of 50--100 km~s$^{-1}$. In NGC~5291--North, we measured a gradient of 100 km~s$^{-1}$ over a distance of 2.4 kpc (8 arcsec), confirming the earlier long--slit observations of  Duc \& Mirabel (1998). We could map the inner velocity field and infer from it that this object is self-gravitating and rotating, although non--circular motions are also clearly observed. The dynamical mass derived form the rotation curve and non-circular motions, and the visible mass, are both of the order of $10^9$ M$_{\sun}$. This may indicate that this object does not contain dark matter inside its HII radius.
 However the uncertainties on both masses remain large. A discrepancy of a few tens of percent between these two masses is still fully consistent with the data. If dark matter is present within the regions delineated by the
ionized gas in the tails, its contribution in mass is not expected to be much higher than a few tens of percent.
\end{enumerate}

Real, self-gravitating objects, with typical masses of $10^9$ M$_{\sun}$, do exist in several systems. It is then justified to consider them as the possible progenitors of dwarf galaxies. 
According to numerical simulations, the formation of $\sim 10^9$ M$_{\sun}$ condensations of matter in the outer regions of tidal tails can be achieved in the extended dark matter halos that are predicted by standard cosmological theories (Navarro et al. 1996), without strong constraints on the galactic encounter parameters (Bournaud et al. 2003). This model accounts for the formation of structures that are similar to the massive TDG progenitors studied in this paper. Yet, the survival of TDGs after their formation should still be studied in detail, since they could be dispersed by star formation, dynamical friction and tidal disruption. Furthermore, whether a significant fraction of dwarf galaxies have a tidal origin is still unknown. Theoretically, their frequency may be determined carrying out realistic numerical simulations of mergers, studying the number of TDGs produced according to the initial parameters and their survival in the face of violent star formation, dynamical friction on dark halos, and tidal disruption. Observationally, it would be worthwhile to try to identify 'old' TDGs of at least 1 Gyr -- as they are then cosmologically significant --, and estimate their numbers in various environments (Duc et al. 2004).

%%%%%%%%%%%%%%%%%%%%%%%%%%%%%%%%%%%%%%%%%%%%%%%%%%%%%%%%%%%%%%%%%%%%%%%%%%%%

\appendix
\section{Details of observed systems}\label{appen}

This appendix details the observed systems and present in particular the
integrated H$\alpha$ emission maps. They have been determined according to the method described in 
Sect.~\ref{analyse}. The location of the velocity gradients studied in Sect.~4 are described
and a few other striking kinematical features presented. We also show velocity fields for systems in which the kinematics is resolved but has not been detailed in the main body of the paper.

\subsection{IC~1182}

This object located in the Hercules cluster is an advanced merger remnant, showing two HI and stellar tails.
A TDG candidate with an apparent total HI mass of $1.8 \times 10^{10}$ M$_{\sun}$
is found at the apparent extremity of the eastern tail (see Dickey 1997; Braine et al. 2001; Iglesias-P\'aramo et al. 2003). 
HII regions are present all along the eastern tail which is seen edge--on (Fig.~\ref{anx1182}).
% A velocity gradient of a few 10 km~s$^{-1}$ has been detected in an HII region, but is not fully resolved.

\begin{figure}
\centering
\caption{Top: H$\alpha$ emission map of IC~1182. A DSS blue image of the same field is shown in the corner. HI contours (Dickey 1997) are superimposed in green. 
Bottom: PV diagram along the slit shown in the map (5 kpc=7''). North-East is to the left of the position axis. 
The asymmetric aspect of this diagram suggests the presence of an internal velocity gradient that is not fully resolved, 
its amplitude but could be as large as 70 km~s$^{-1}$. The orientation of the slit corresponds to the largest velocity gradient.}
\label{anx1182}
\end{figure}

\subsection{Arp~105 (NGC 3561)}

The interacting system Arp~105, located in the cluster Abell 1185, has been extensively studied by Duc \& Mirabel (1994) and Duc et al. (1997). It consists of a spiral galaxy tidally interacting with an elliptical galaxy, one major tidal tail heading to the North, and one counter-tail to the South, passing in front of the elliptical. Both tails that are seen nearly edge-on show at their tip HI condensations with apparent masses of respectively $6.5 \times 10^{9}$ M$_{\sun}$ and $0.5 \times 10^{9}$ M$_{\sun}$. Its relative high oxygen abundance makes it a suitable TDG candidate. A possible velocity gradient that is not fully resolved has been detected in the Southern TDG candidate (see Sect.~4.1 and Fig.~A.2)

%We have detected several HII regions in the southern tail (Fig.~\ref{anx105}), enabling to study the large-scale structure
% of this tail. A velocity gradient is also detected in the southern TDG candidate, it is not fully resolved, but the upper limit
% to its actual value is 100 km~s$^{-1}$. The H$\alpha$ emission of the northern tail has not been studied yet.

\begin{figure}
\centering

\caption{Top: H$\alpha$ emission map of Arp~105. A DSS blue image of the same field is shown in the corner, and HI contours are superimposed (Duc \& Mirabel 1998). Three main objects can be identified, from north to south: the spiral galaxy, the elliptical one, and the TDG candidate. Bottom: PV diagram of the southern TDG along the slit shown in the map (5 kpc=8''). South-East is to the left of the position axis. The asymmetric aspect of this diagram suggests the presence of an inner velocity gradient, that is not resolved but may reach 100 km~s$^{-1}$ over 2 to 3 kpc. The orientation of the slit corresponds to the largest velocity gradient. }
\label{anx105}
\end{figure}

\subsection{Arp~242 (NGC 4676, The Mice)}
Arp~242 consists of two merging spiral galaxies. Each one has developed a long HI and stellar tail 
(Hibbard \& van Gorkom 1996). An HI cloud with an apparent mass of $2.2 \times 10^{9}$ M$_{\sun}$ is observed
at the tip of the northern tail which is seen almost perfectly edge--on (Hibbard \& Barnes 2004).
 About $10^{8}$ M$_{\sun}$ of molecular gas was detected towards this TDG candidate (Braine et al. 2001).
We have found many HII regions distributed along this tail, the main one corresponding to the TDG candidate (Fig.~\ref{anx242}).

\begin{figure}
\centering
\caption{H$\alpha$ emission map of Arp~242 with HI contours superimposed in green (Hibbard \& van Gorkom 1996).
 A DSS blue image of the whole system is shown in the corner. The two 
straight lines indicate the band along which the PV diagram of the 
northern 
tail shown in Fig.~\ref{PV242} has been established.} \label{anx242}
\end{figure}

\subsection{NGC~5291}

NGC~5291 is located in the cluster of galaxies Abell 3574. It consists of a perturbed lenticular
galaxy and a disrupted spiral (''The Seashell''), surrounded by a giant HI 
ring, which
extends over almost 200 kpc and contains about $5 \times 10^{10}$ M$_{\sun}$ of atomic
hydrogen (Malphrus et al. 1997). The origin of the ring which has a mean velocity similar 
to the systemic velocity of the lenticular is still unknown. Although it was probably not shaped
by tidal forces, it most likely results from a galaxy--galaxy collision. Duc \& Mirabel (1998)
identified along the ring numerous HII regions associated with HI condensations with masses
of 0.1 -- 2.5 $\times 10^{10}$ M$_{\sun}$, that we have also observed (Figs.~\ref{anx5291N}, \ref{anx5291C} and \ref{anx5291S}). An inner velocity gradient of the northern TDG candidate has been resolved (see Sect.~4.2). In the central regions, where Duc \& Mirabel (1998) had already found evidence of counter-rotating structures, we have detected kinematical decouplings (see Fig.~\ref{anx5291C}).

\begin{figure*}
%\sidecaption
%\includegraphics[width=12cm]{N5291NHA.eps}
\caption{ Left: H$\alpha$ emission map of the northern region of 
NGC~5291 with HI contours superimposed in green (Malphrus et al. 1997). A DSS blue image of the system is shown in the corner. The ``seashell'' galaxy is seen at its bottom. The line indicates the slit along which the velocity curve of the TDG candidate (Fig.~\ref{PV5291}) has been derived. -- Right: H$\alpha$ velocity field associated with the same region. Velocities indicated in the colorbar are in km~s$^{-1}$.}
\label{anx5291N}
\end{figure*}

\begin{figure}
\centering
\caption{ Top: H$\alpha$ emission map of the central region of NGC~5291 
with HI contours superimposed in green (Malphrus et al. 1997). A DSS blue image of the system is shown in the corner.
 -- Center: PV diagram corresponding to the band (a) on the map above (2 kpc=6.5''). North is to the left of the position 
axis. Note a dedoubling of the kinematical features around the brightest HII region, in the central regions of the parent galaxy. Duc \& Mirabel (1998) had found evidences of counter-rotation is this region -- Bottom: PV diagram corresponding to band (b). East is to the left of the position axis; the dashed line indicates the mean velocity of most HII regions. However, the brightest HII region is largely decoupled from this mean trend.}
\label{anx5291C}
\end{figure}

\begin{figure*}
%\sidecaption
%\includegraphics[width=12cm]{N5291SHA.eps}
\caption{Top: H$\alpha$ emission map of the southern region of NGC~5291 with HI contours superimposed in
 green (Malphrus et al. 1997). A DSS blue image of the system is shown in 
the corner. -- Bottom: PV diagram of 
%the velocity gradient observed in 
one of the TDG candidates along the slit shown in the map (East is to 
the left; 2 kpc=6.5''). -- Right: H$\alpha$ velocity field associated 
with the same region. Velocities indicated in the colorbar are in km~s$^{-1}$.}
\label{anx5291S}
\end{figure*}

\subsection{Arp~243 (NGC 2623)}
Most of the HI component in the advanced merger Arp~243 is located outside 
the stellar main body (Hibbard \& Yun 1996). Whereas the north-eastern tidal tail exhibits many HII regions (Fig.~\ref{anx243}), the western tail has only one giant HII region, located not far from the HI peak, and which could be a TDG candidate. This object shows a small inner velocity gradient whose amplitude is approximatively 40 km~s$^{-1}$. Its mean radial velocity is given in Table~\ref{velos}. Because of the compactness of the H$\alpha$ emission, we were unable to study the global kinematics of the tidal, which is anyway not observed edge--on but almost face--on.

\begin{figure}
\centering
\caption{H$\alpha$ emission map of a northern region of Arp~243 with HI contours superimposed in green (Hibbard \& Yun 1996). A DSS blue image of the same field is shown in the corner. The giant HII region associated with the western tail has been circled. The H$\alpha$ velocity field of the northern tail is shown below.}
\label{anx243}
\end{figure}

\subsection{Arp~244 (NGC 4038/9, The Antennae)}
The two spectacular stellar and gaseous tails that emerge from the merging 
spiral galaxies NGC 4038/9 are well known. 
The massive accumulation of stellar and gaseous matter (containing about $3 \times 10^{9}$ M$_{\sun}$ of HI)
 in the outer regions of the southern tail was the first TDG candidate (Schweizer 1978; Mirabel et al. 1992).
It was later studied in detail by  Hibbard et al. (2001) who failed to detect in it a kinematic signature of a 
self-gravitating condensation. Braine et al. (2001) reported faint CO emission towards that direction.
We present in Fig.~\ref{anx244} an H$\alpha$ map of the region (the first one to our knowledge).
Quite surprisingly, the H$\alpha$ emission is restricted to at least 8 extremely compact HII regions located close to the HI emission peak. These regions are too sparsely distributed to enable any study of their global and inner kinematics. Their radial velocities are given in Table~\ref{velos}. We have only selected the HII region candidates associated with an emission line detected above 2$\sigma$ over at least 6 channels, and that are not associated with stars. Other regions seen in Fig.~A.8 are believed to be either noise (no emission line detected over more than 3 channels), stars, or artefacts of bright stars in the interferometer.

% However, the presence of HII regions near the HI peak, while no HII regions are detected anywhere else in the 
%observed region of the tail, probably indicates that there is a genuine concentration of matter, and that the observed HI 
%peak is not only the result of a projection effect, since star formation occurs only in this region.

\begin{table*}
\centering
\begin{tabular}{lccc}
\hline
\hline
Object & RA (J2000) & DEC (J2000) & Radial velocity (km~s$^{-1}$) \\
\hline
Arp244--1 & 12h 01m 35.8s & -19$^{\circ}$ 01' 13'' & 1763 \\
Arp244--2 & 12h 01m 27.8s & -19$^{\circ}$ 00' 49'' & 1770 \\
Arp244--3 & 12h 01m 27.0s & -19$^{\circ}$ 00' 54'' & 1900 \\
Arp244--4 & 12h 01m 25.0s & -19$^{\circ}$ 00' 56'' & 1903 \\
Arp244--5 & 12h 01m 27.4s & -19$^{\circ}$ 00' 20'' & 1856 \\
Arp244--6 & 12h 01m 21.7s & -19$^{\circ}$ 00' 13'' & 1837 \\
Arp244--7 & 12h 01m 22.6s & -18$^{\circ}$ 59' 21'' & 1880 \\
Arp244--8 & 12h 01m 23.1s & -19$^{\circ}$ 02' 03'' & 1822 \\
Arp215    & 09h 40m 40.9s & +40$^{\circ}$ 08' 12'' & 2520 \\
Arp243    & 08h 38m 19.2s & +25$^{\circ}$ 44' 06'' & 5530 \\
\hline
\end{tabular}
\caption{Individual coordinates and radial velocities of the compact HII region candidates observed in Arp~244 (see the number attributed to each region in Fig.~\ref{anx244}), Arp~215, and Arp~243.}\label{velos}
\end{table*}

\begin{figure}
\centering
\caption{H$\alpha$ emission map of a northern region of Arp~244 with HI contours superimposed in  green (Hibbard et al. 2001). A DSS red image of the system is shown in the corner, and the field of the H$\alpha$ map delimited by the red rectangle. We have circled HII regions associated with an H$\alpha$ line visible in the data cube, to distinguish these regions from noise (not associated with a typical emission line) and artefacts due to stars in the Fabry-P\'erot interferometer.}
\label{anx244}
\end{figure}

\subsection{Arp~215}

The peculiar system Arp~215 has been studied in detail by Smith (1994) and Smith et al. (1999). It shows a 
stellar tail extending to the east, and, in the opposite direction, an extended HI plume. A compact  HII region associated with the HI plume is detected and its radial velocity is given in Table~\ref{velos}, but most of the  the H$\alpha$ emission is restricted to smaller radii, inside the central disk galaxy (Fig.~\ref{anx215}).

\begin{figure}
\centering
\caption{H$\alpha$ emission map of a northern region of Arp~215 with HI contours superimposed in green (Smith 1994). A DSS blue image of this object is shown in the corner. The HII region associated with the HI plume is circled.}
\label{anx215}
\end{figure}

\subsection{Arp~245 (NGC~2992/3)}

The interacting pair of galaxies Arp~245 has been studied by Duc et al. (2000). The tidal tail associated with the southern spiral is highly curved and shows no obvious TDG candidate. The northern tail, seen close to edge-on, exhibits a condensation of stars and HI at its extremity, as massive as 10$^9$ M$_{\sun}$. Our observations confirm that the  emission of the northern system is restricted to the parent galaxy (NGC~2992) and the TDG candidate, as found by Duc et al. (2000). The absence of H$\alpha$ emission along the tail does not allow us to study the large-scale kinematics of the tail, so we do not obtain any constraints on its geometry. The nature and inner dynamics of the TDG candidate will be analyzed using the present HII data together with HI high resolution observations (Brinks et al. 2004).

\begin{figure}
\centering
\caption{H$\alpha$ emission map of a northern region of Arp~245 with HI 
contours superimposed in green (Duc et al. 2000). A DSS blue image of this object is shown in the inset, and the field of the H$\alpha$ map is delimited by the red rectangle.}
\label{anx245}
\end{figure}

%%%%%%%%%%%%%%%%%%%%%%%%%%%%%%%%%%%%%%%%%%%%%%%%%%%%%%%%%%%%%%%%%%%%%%%%%%%%%%

\begin{acknowledgements}
We are grateful to the referee, Almudena Zurita, for her very careful reading of the manuscript and useful suggestions to clarify the text and the figures. We thank the CFHT's support astronomer Pierre Martin and Chantal Balkowski, Olivier Boissin and Jacques Boulesteix for their help during the ESO observations. We are grateful to Jacques Boulesteix for maintaining the ADHOC software, and to John Hibbard, Beverly Smith, Elias Brinks, John Dickey and Benjamin Malphrus for providing us with the HI maps shown in the paper. The N-body simulations in this work were computed on the Fujitsu NEC-SX5 of the CNRS computing center, at IDRIS. This research made use of the NASA/IPAC Extragalactic Database (NED) which is operated by the Jet Propulsion Laboratory, California Institute of Technology, under contract with the National Aeronautics and Space Administration, and of the Digitized Sky Survey. The Digitized sky Survey was produced at the Space Telescope Science Institute under U.S. Government grant NAG W-2166. The images of these surveys are based on photographic data obtained using the Oschin Schmidt Telescope on Palomar Mountain and the UK Schmidt Telescope. The plates were processed into the present compressed digital form with the permission of these institutions.
\end{acknowledgements}

\end{document}